\documentclass[twocolumn,showpacs,preprintnumbers,amsmath,amssymb,prl,superscriptaddress]{revtex4-1}
\usepackage[T1]{fontenc}
\usepackage[latin1]{inputenc}
\usepackage{graphicx}

\usepackage{amsmath,amssymb}
\usepackage{xcolor}
\newcommand{\etal}{\mbox{\emph{et al.}\hspace{-1pt}}~}

\newcommand{\ev}{\mathrm{eV}}

\newcommand{\angstrom}{\textup{\AA}}
\newcommand{\nm}{\mathrm{nm}}
\newcommand{\cm}{\mathrm{cm}}

\newcommand{\Eqref}[1]{Eq.~(\ref{#1})}
\newcommand{\Figref}[1]{Fig.~\ref{#1}}

\makeatother

\newcommand{\siesta}{\textsc{Siesta}}
\newcommand{\tsiesta}{\textsc{TranSiesta}}
\newcommand{\tbtrans}{\textsc{TBtrans}}
\newcommand{\phtrans}{\textsc{PHtrans}}
\newcommand{\sisl}{\textsc{sisl}}

\expandafter\let\csname equation*\endcsname\relax
\expandafter\let\csname endequation*\endcsname\relax
\usepackage{amsmath,newtxmath,amssymb}

\newif\ifchanged
\changedtrue
\changedfalse

\newcommand{\change}[1]{\ifchanged\textcolor{red}{#1}\else#1\fi}

\begin{document}

\title{Multi-scale approach to first-principles electron transport beyond 100 nm}

\author{Gaetano Calogero}
\affiliation{Dept. of Physics, Technical University of Denmark, DK-2800 Kongens Lyngby, Denmark}
\affiliation{Center for Nanostructured Graphene (CNG), DK-2800 Kongens Lyngby, Denmark}

\author{Nick R. Papior}
\affiliation{Computing Center, Technical University of Denmark, DK-2800 Kongens Lyngby, Denmark}
\affiliation{Center for Nanostructured Graphene (CNG), DK-2800 Kongens Lyngby, Denmark}

\author{Mohammad Koleini}
\affiliation{Dept. of Physics, Technical University of Denmark, DK-2800 Kongens Lyngby, Denmark}
\affiliation{Center for Nanostructured Graphene (CNG), DK-2800 Kongens Lyngby, Denmark}

\author{Matthew Helmi Leth Larsen}
\affiliation{Dept. of Physics, Technical University of Denmark, DK-2800 Kongens Lyngby, Denmark}

\author{Mads Brandbyge}
\affiliation{Dept. of Physics, Technical University of Denmark, DK-2800 Kongens Lyngby, Denmark}
\affiliation{Center for Nanostructured Graphene (CNG), DK-2800 Kongens Lyngby, Denmark}

\email{tanocalogero92@gmail.com, mabr@dtu.dk}

\begin{abstract}
  Multi-scale computational approaches are important for studies of novel, low-dimensional electronic devices since they are able to capture the different length-scales involved in the device operation, and at the same time describe critical parts such as surfaces, defects, interfaces, gates, and applied bias, on a atomistic, quantum-chemical level.
		Here we present a multi-scale method which enables calculations of electronic currents in two-dimensional devices larger than 100 nm$^2$, where multiple perturbed regions described by density functional theory (DFT) are embedded into an extended unperturbed region described by a DFT-parametrized tight-binding model. We explain the details of the method, provide examples, and point out the main challenges regarding its practical implementation. Finally we apply it to study current propagation in pristine, defected and nanoporous graphene devices, injected by chemically accurate contacts simulating scanning tunneling microscopy probes.
\end{abstract}

\maketitle

%%%MAIN TEXT%%%%

%%%%%%%%%%%%%%%%%%%%%
\section{Introduction}
Developing high-performance computational strategies to simulate electronic devices is a fundamental asset for prototype design and research planning in basically any technological context.\cite{Hofer2018, Marzari2016,Thijssen2007} Being able to model nanometer-scale devices with atomic resolution has become particularly crucial for e.g.\ novel low-dimensional materials, molecular junctions, or, generally, ballistic quantum systems which have appeared on the electronics horizon in the last decades.\cite{Aprojanz2018, Shulaker2013,Ferrari2015,Novoselov2016,Geim2013,Martinez-Blanco2015,Moreno2018,Jia2016,Cao2018}
Quantum-chemical details are often critical for describing local electronic structure, chemical contacts (e.g. electrodes\cite{Kretz2018,Palacios2008}) and electrostatics in realistic nano-electronic devices. This is especially clear in a context of one-atom-thick two-dimensional (2D) devices,\cite{Fiori2014,Novoselov2016,Iannaccone2018} where details on the atomic scale govern the electronic behavior.\cite{Schedin2007, Wehling2008, Chang2012, Caridad2018, talbotNPG} 
These details can be accessed using various {\it ab-initio} methods,\cite{Szabo1989,Levine2014, Thijssen2007} in particular density functional theory (DFT),\cite{Kohn1965,Hohenberg1964} which can be used to simulate up to thousands of atoms.\cite{Goedecker1999, Mohr2015, Lin2014}
This is typically enough to simulate isolated portions of a realistic device, such as bulk regions, interfaces or locally perturbed areas, but it is not suitable to simulate all the different length-scales involved in the operation of realistic devices. 

A promising solution which is rapidly growing along many lines of research is to perform hybrid multi-scale simulations.\cite{Hofer2018,Hofer2015,Troisi2006,Kubar2010,Zhang2018,Shen2016,Fiori2013,AufDerMaur2011,Fediai2016,Fediai2016b} 
In this context an interesting approach is to treat crucial parts of a relatively large device using full quantum-chemical detail, while reducing the number of degrees of freedom to the bare essentials elsewhere.\cite{Hofer2018,Hofer2015,Troisi2006,Kubar2010,Zhang2018,Shen2016,Ng2008,Rapacioli2011} A very popular example, coming from a biomolecular context, is the quantum mechanics/molecular mechanics (QM/MM) technique,\cite{Warshel1976} which has also been generalized to study solid-state surfaces and interfaces.\cite{Hofer2015} 
The key problem of multi-scale approaches is to partition the system into a number of subregions and, most importantly, to ensure a smooth, physically sound, transition among them.\cite{Hofer2018}

Electronic structure methods employing linear combinations of atomic orbitals (LCAO) provides an intuitive route for elaborating multi-scale approaches analogous to QM/MM for simulating electron transport, as the accuracy and scalability of LCAO-based approaches can be ``tuned'' via clever approximations of the Hamiltonian.
One common case is the Density Functional Tight-Binding (DFTB), which is based on a Taylor series expansion of the Kohn-Sham DFT total energy.\cite{Elstner1998, Elstner2014} Once a reliable element or compound-specific parametrization has been obtained, either comparing to experiments or higher-level {\it ab-initio} calculations, DFTB can be two to three orders of magnitude faster than DFT,\cite{Gaus2013} overall enabling simulations with several thousand atoms without need for any massive parallelization.\cite{Elstner2014}
Another useful approximation is the Wannier Tight-Binding (WTB), where maximally localized Wannier functions (MLWF) are constructed {\it ad-hoc} to reproduce electronic bands from plane-wave DFT within optimized energy windows.\cite{Marzari1997, Marzari2012, Bruzzone2014, Marin2018,Lv2015,Pizzi2016} Once a system-specific parameterization is found, and good initial trial orbitals are defined,\cite{Gresch2018} one can use WTB to access samples containing hundreds of thousands of atoms.\cite{Rudenko2015} 
The latter two tight-binding (TB) like approximations can be ultimately simplified by reducing the number of orbitals per atom to a minimum, and limiting the interaction range to only a few nearest neighbors in the lattice, fitting parameters by hand to experiments or {\it ab-initio} calculations.\cite{Martin2004, Ashcroft, Liu2015, Beconcini2016, Calogero2018}
Multi-scale approaches for electron transport based on these models have been proposed and include combinations of Classical Molecular Dynamics (CMD) and DFT with Langevin Dynamics,\cite{Troisi2006} combinations of CMD, DFTB, and wave function propagation,\cite{Kubar2010} or combinations of TB models with patched Green's function techniques.\cite{Settnes2015}

%------------
% Our goal
In this article we present a multi-scale scheme based on seamless integration of a number of \emph{perturbed} regions described by DFT and an \emph{unperturbed} extended region represented by a simpler representative LCAO model, namely TB. We parametrize the TB Hamiltonian directly from DFT, such that the resulting multi-scale model is defined \emph{without any fitting parameter or by hand adjustments}. The obtained models can be combined with the non-equilibrium Green's function (NEGF) formalism\cite{TaGuWa2001,Brandbyge2002, Datta2000, Papior2017} to enable current simulations of over $100 \nm\, \times 100\, \nm$ large devices with local quantum-chemical detail.
The manuscript is divided into two main sections.
In the first section we develop the general formalism, didactically providing concrete examples based on simple graphene two-electrode devices. 
In the second section we apply the multi-scale approach for imaging real-space ``far-field'' currents (i.e. far from the source) in pristine, defected and nanoporous graphene, injected by chemically accurate contacts simulating Scanning Tunnel Microscopy (STM) probes.

%%%%%%%%%%%%%%%%%%%%%
\section{Method}

\subsection{TB models from orbital-projected DFT}
\label{sec:tb-projection}

\begin{figure}[t]
	\includegraphics[width=1.0\columnwidth]{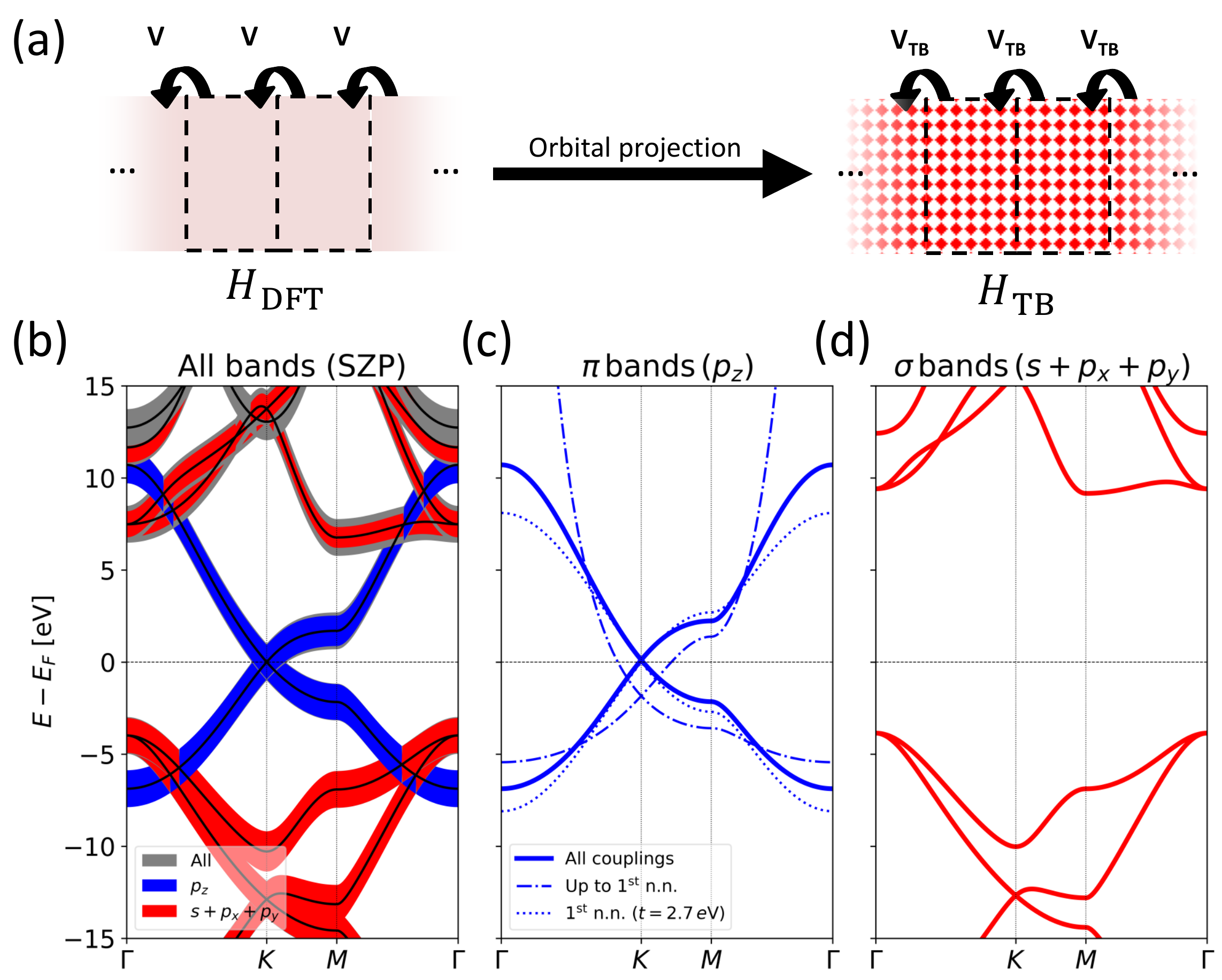}
	\caption{
		(a) Schematic illustration of the orbital-projection method to create TB models out of DFT Hamiltonians. The red/white pattern indicates regions defined by a subset of the original DFT basis.
		(b) Orbital-resolved graphene bandstructure from a DFT calculation with a single-$\zeta$ polarized (SZP) basis set (9 orbitals per C atom). Contributions from $p_z$ ($s + p_x + p_y$) orbitals are highlighted in blue (red). Orbital weights are normalized to the total contribution given by all orbitals (grey).
		(c) Graphene $\pi$ bands from a 1-orbital TB model generated by projecting
		the DFT Hamiltonian onto all $p_z$ orbitals in the system (solid), in
		comparison with those from a standard $1^{\mathrm{st}}$ nearest neighbor
		orthogonal TB model with hopping $t=-2.7\,\ev$ (dotted). Dashed-dotted is
		DFT subset \emph{and} limiting to $1^{\mathrm{st}}$ couplings.
		(d) Graphene $\sigma$ bands from a 3-orbital TB model generated by
		projecting the DFT Hamiltonian onto all $s$, $p_x$ and $p_y$ orbitals in
		the system.  }
	\label{fig:grbands}
\end{figure}

Our starting point is a pristine structure of a periodically repeated unit cell, illustrated in \Figref{fig:grbands}, and a DFT Hamiltonian describing this system with a localized basis. We will here consider a LCAO basis for the DFT calculation, $\phi_{\alpha}$, but it could equally well be a basis obtained from e.g. maximally localized Wannier states. We will from this construct a smaller TB-like basis, which only describe the bands in a region around the Fermi energy, $E_F$, set via a projection $P$,
\begin{equation}
\{\bar\phi_\alpha \} = P\{\phi_\alpha \}\,.
\end{equation}
In the simplest case, used here, this involves selecting a particular subset of the original DFT basis functions, however one may imagine more involved projections.
Our prime example is shown in \Figref{fig:grbands}b, where we consider the bandstructure of graphene obtained with DFT and a LCAO single-$\zeta$ plus polarization basis
(i.e. a single orbital for each $s$, $p$ states and polarization $d$ orbitals).
A TB-like model can be readily constructed by projecting on to the $p_z$ orbitals of the DFT LCAO
Hamiltonian (and overlap), by extracting all rows and columns associated to $p_z$
orbitals.  This cheaper model is sufficient to capture the $\pi$ bands of graphene
(\Figref{fig:grbands}c). Likewise, the $\sigma$ bands further away from the Dirac point
energy can be captured using a TB model parametrized from $s$, $p_x$ and $p_y$ DFT
orbitals (\Figref{fig:grbands}d).  Importantly, as highlighted (dashed lines) in
\Figref{fig:grbands}c, a partial $p_z$-projection where only couplings among nearest
neighbors are retained results in significant band-misalignment and rescaling. Analogous
but less dramatic deviations are observed when using a standard nearest neighbor model
with orthogonal basis and hopping $t=2.7\,\ev$.

%%%%%%%%%%%%%%%%%%%%%
\subsection{Multi-scale approach}
In the following we omit energy and $\mathbf k$ dependence, highlighting it only where
necessary.

The multi-scale method presented in this work is based on the Non-Equilibrium Green's
Function (NEGF) transport formalism.\cite{TaGuWa2001,Brandbyge2002, Datta2000, Papior2017}  In the NEGF
framework transmission between any two leads $i$ and $j$ of a $N$-electrode device with
Hamiltonian $\mathbf H$ and overlap $\mathbf S$ is given by
\begin{gather}
\mathcal T_{i j} = \mathrm{Tr} \left[ \mathbf G_D \boldsymbol\Gamma_{i} \mathbf G^{\dagger}_D \boldsymbol\Gamma_{j} \right] \label{eq:t} \\
\boldsymbol\Gamma_j = i \left( \boldsymbol\Sigma_j - \boldsymbol\Sigma^\dagger_j \right) \label{eq:gams}
\end{gather}
where the device Green's function $\mathbf G_D$ is given by
\begin{equation}
\label{eq:green}
\mathbf G_D = \left[ \mathbf S\,(E + i\eta) - \mathbf H - \sum^N_{i}\boldsymbol\Sigma_i \right]^{-1}
\end{equation}
and $\boldsymbol\Sigma$ are the so-called \emph{self-energies} of the semi-infinite electrodes.

It is the self-energies that will play a pivoting role in connecting TB and
DFT models. The self-energy is nothing but the effect of degrees of freedom not accounted explicitly for in the equation but have been eliminated as can be done for variables in linear equations. The self-energies can represent the semi-infinite electrodes as well as a finite number of degrees of freedom in the case of a finite region. 
In the following we show that it is in practice possible to locally replace small
perturbed regions of a large TB device with DFT-precision models by simply including one
or more ``special'' self-energies in the sum of \Eqref{eq:green}. 

\subsubsection{Self-energy for a partitioned system}

To understand the construction of these special self-energies it is instructive to recall
the definition of self-energy connecting two subregions of a generic binary system. We do
this by simply generalizing the derivation of self-energy for a simple system divided into two parts.\cite{PapiorThesis}

\begin{figure*}[t]
	\includegraphics[height=0.28\textheight]{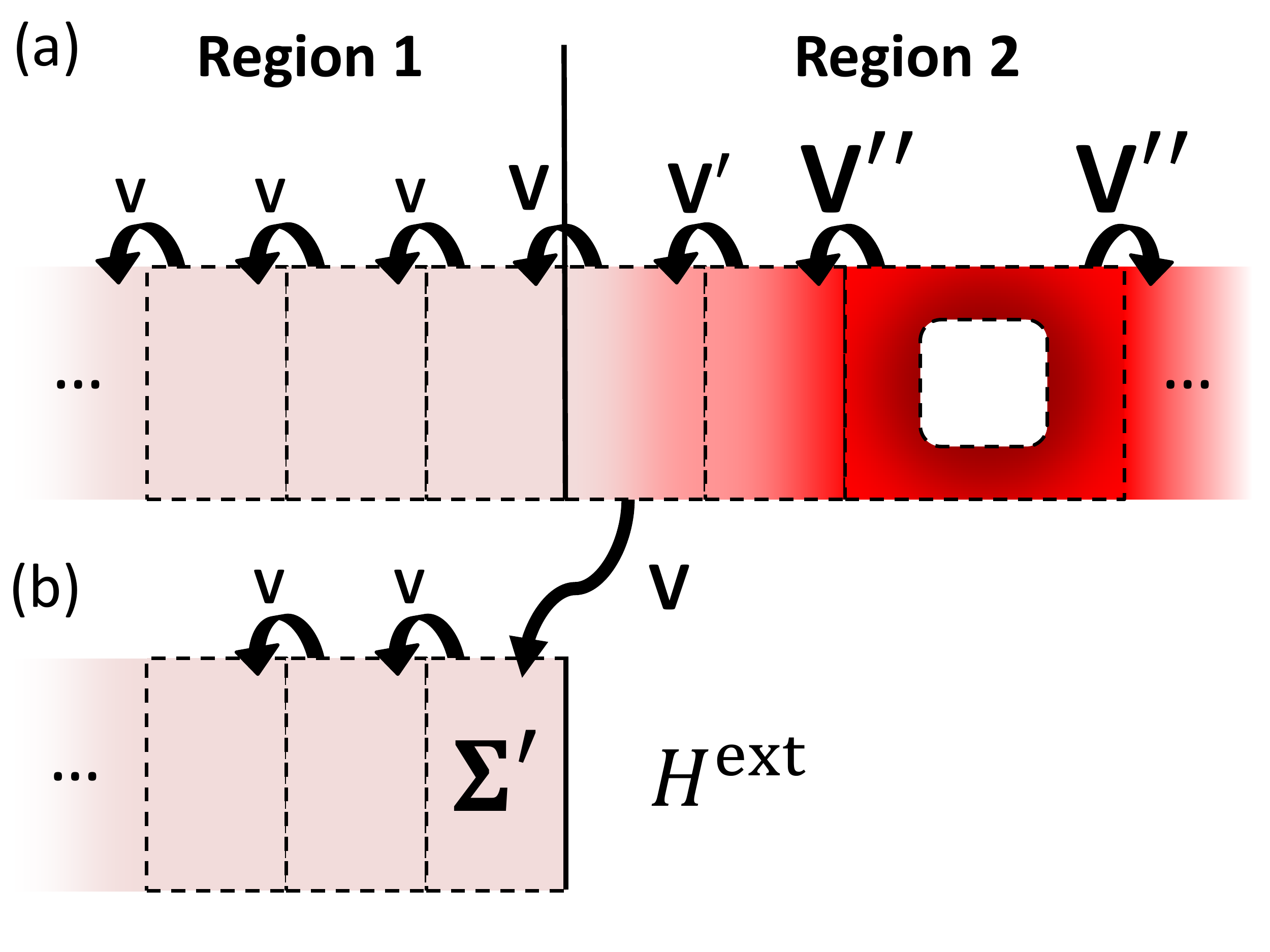}
	\includegraphics[height=0.28\textheight]{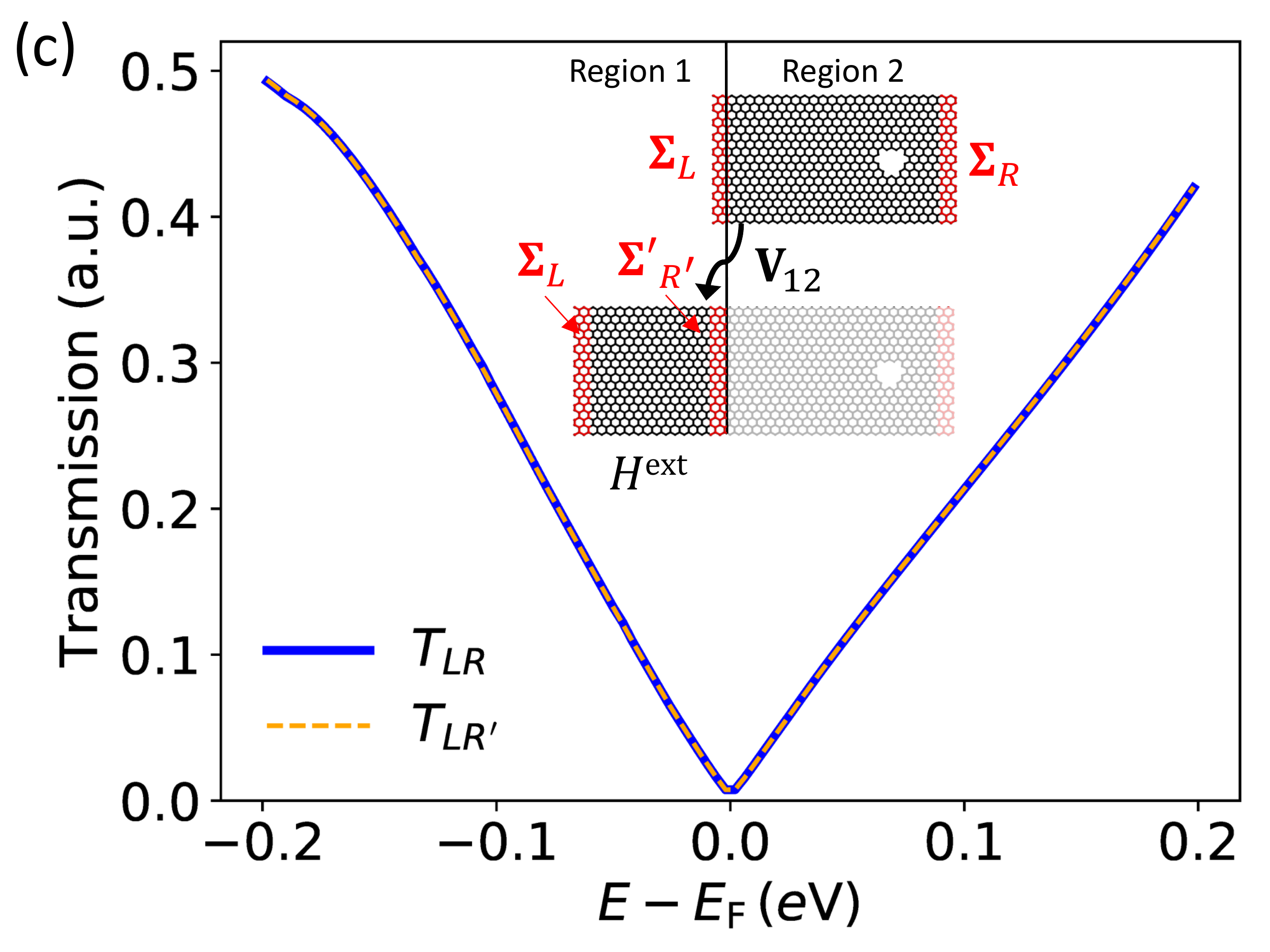}
	\caption{(Color online)
		(a) Illustration of a system divided into a pristine region 1, consisting of a repeated unit cell, and a perturbed region 2 having a hole in its structure. The potential $\mathbf V''$ (red) in proximity of the hole is screened far from it (pink), saturating to a constant value $\mathbf V$.
		(b) Illustration of an unperturbed system where the hole is substituted by a self-energy $\boldsymbol\Sigma'$.
		(c)
		Comparison between transmission across a DFT two-electrode graphene device with a
		hole ($\mathcal T_{LR}$) and a DFT device with pristine scattering region and a
		self-energy $\boldsymbol\Sigma'_R$ substituting the hole ($\mathcal T_{LR'}$). 
		Geometries are shown in the inset with the electrodes highlighted in red. The opaque area on the right side of the lower panel indicates the region which is replaced by $\boldsymbol\Sigma'$.
	}
	\label{fig:se}
\end{figure*}

Let us consider the system in \Figref{fig:se}a, with Hamiltonian $\mathbf H$, and overlap
$\mathbf S \equiv \mathbf 1$ (we use an orthogonal basis without loss of generality),
\begin{equation}
\mathbf H + \mathbf V
= \left[
\begin{matrix}
\mathbf H_{1,1} & 0
\\
0 & \mathbf H_{2,2}
\end{matrix}
\right] + \left[
\begin{matrix}
0 & \mathbf V_{1,2}
\\
\mathbf V_{2,1} & 0
\end{matrix}
\right],\,.
\label{eq:hv}
\end{equation}
Here we have explicitly indicated with $\mathbf V$ the coupling of region 1
($\mathbf H_{1,1}$) to the neighboring region 2 ($\mathbf H_{2,2}$).
In the following we are only interested in the Green's function in region $\mathbf
G_{1,1}$, and thus a derivation of this matrix is required.
The full Green's function is
\begin{equation}
\left[ z\mathbf 1 - \mathbf H - \mathbf V \right] \mathbf G(E) = \mathbf 1,
\end{equation}
with $z=E + i\eta$ and $\eta \to0^+$.
By using the Dyson equation
\begin{equation}\label{eq:dyson}
\mathbf G = \mathbf G^0 + \mathbf G^0 \mathbf V \mathbf G,
\end{equation}
we can express the Green's function in only one region
\begin{align}
\label{eq:g11}
\mathbf G_{1,1} &= \mathbf G^0_{1,1} + \mathbf G^0_{1,1} \mathbf V_{1,2} \mathbf
G_{2,1},
\\
% \GC{Note that this is true only because $G^0_{1,2}=G^0_{2,1}=0$.}
\mathbf G^0_{1,1}&=[z\mathbf 1 - \mathbf H_{1,1}]^{-1},
\\
\label{eq:g21}
\mathbf G_{2,1} &= \mathbf G^0_{2,2} \mathbf V_{2,1} \mathbf G_{1,1},
\end{align}
and thus
\begin{align}
\mathbf G_{1,1} &= \mathbf G^0_{1,1} + \mathbf G^0_{1,1} \mathbf V_{1,2} \mathbf G^0_{2,2} \mathbf V_{2,1} \mathbf G_{1,1} \\
&= \left[ z \mathbf I - \mathbf H_{1,1} - \mathbf V_{1,2} \mathbf G^0_{2,2} \mathbf V_{2,1} \right]^{-1} \\
&= \left[ z \mathbf I - \mathbf H_{1,1} - \boldsymbol\Sigma' (E) \right]^{-1}\,,
\end{align}
Here the term
\begin{equation}
\label{eq:sigmap}
\boldsymbol\Sigma'(E)= \mathbf V_{1,2} \, \mathbf G^0_{2,2} (E) \, \mathbf V_{2,1}\,,
\end{equation}
is the self-energy describing how region 1 is perturbed by the coupling to the degrees of freedom in region 2. 
In case of a non-orthogonal basis set ($\mathbf S \nequiv \mathbf 1$) \Eqref{eq:sigmap} simply becomes
\begin{equation}
\label{eq:sigmapS}
\boldsymbol\Sigma'(E)= \mathbf (\mathbf V_{1,2} - E \mathbf S_{1,2}) \, \mathbf G^0_{2,2} (E) \, (\mathbf V_{2,1} - E \mathbf S_{2,1})\,.
\end{equation}
The real part of the self-energy is an energy renomarlization/shift, while the, possibly finite, imaginary part corresponds to a finite life-time  or broadening of the energy levels in region 1. Additionally the self-energy $\boldsymbol\Sigma'$ is
independent of $\mathbf H_{1,1}$ and may thus be used in any other system
$\mathbf H_{\rm ext}$ so long as the coupling matrix is unchanged, i.e. $\mathbf V_{1,2} = \mathbf V_{{\rm ext},2}$.

In \Figref{fig:se}a, we see the above exemplified using an extended system with a hole
perturbing the local potential (dark colors). After two layers the perturbation goes to zero and the coupling between neighbouring cells turns to the bulk value($\mathbf V$). At this point one
can calculate $\boldsymbol\Sigma'$ for region 2, insert it into an external \emph{pristine} system (equal to region 1 in \Figref{fig:se}b) and, in turn, reproduce the correct Green's function in region 1, as though the new system was connected with the hole.

\subsubsection{Example: graphene device with DFT-DFT connection}

In the following we illustrate with a concrete example where the
two models, connected by $\boldsymbol\Sigma'$, are constructed using the same method (here DFT with a SZ basis set, i.e.\, 4 orbitals per C atom), then effects of the perturbation can be propagated from one model to the other with 100\% accuracy. Consider the basic DFT two-electrode graphene device illustrated in
\Figref{fig:se}c (upper inset). This has periodic boundary conditions along $y$,
semi-infinite electrodes $L$ and $R$ along $\pm x$ and
a hole in the scattering region. Total transmission $\mathcal T_{LR}$ across the device
can easily be computed by using \Eqref{eq:t} and is plotted in \Figref{fig:se}c.
One can choose to split this device into two sections, a bulk part (region 1) and a
perturbed region 2 (hole). Since the potential of the hole is fully screened in region 1, we have that the coupling between $\mathbf V_{2,1}$ equals the bulk coupling, $\mathbf V$.
Hence one can construct the self-energy $\boldsymbol\Sigma'_{R'}$ which contains the effects of region 2 propagated into region 1 using \Eqref{eq:sigmap}. This can then be
incorporated as a new electrode on the right side of an external DFT device which has no
holes in the scattering region (lower inset) and verify that
$\mathcal T_{LR'} (E) = \mathcal T_{LR} (E)$ for all $E$ within the numerical accuracy.

\begin{figure*}[t]
	\includegraphics[height=0.32\textheight]{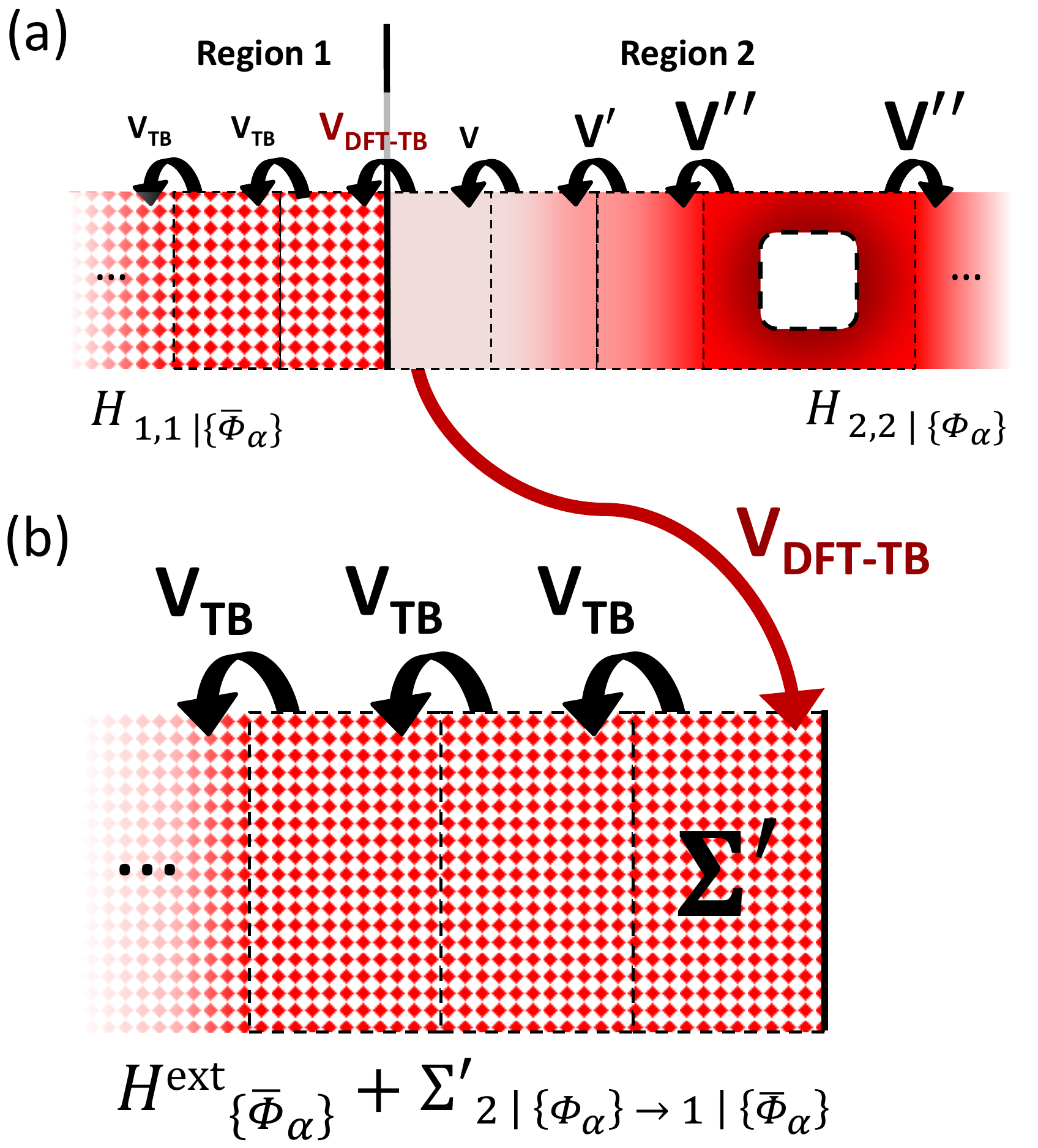}
	\includegraphics[height=0.32\textheight]{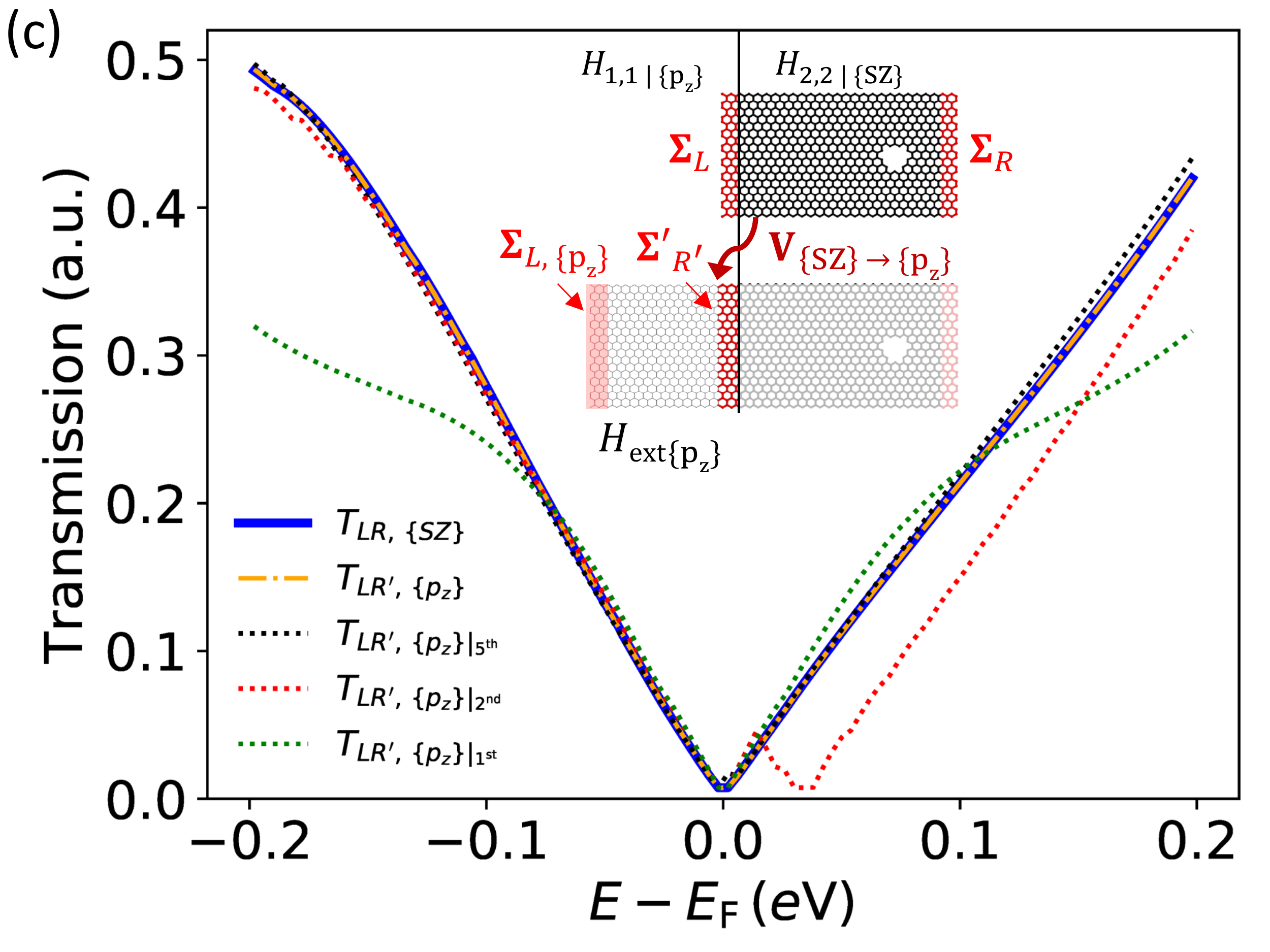}
	\caption{ (a) Generic DFT model for a binary system. Region 1 is defined using a subset
		$P\{\phi_{\alpha}\}=\{\bar{\phi}_{\alpha}\}$ of orbitals from the original DFT basis set
		$\{\phi_{\alpha}\}$. Region 2 has a hole in its structure and is defined using
		$\{\phi_{\alpha}\}$. The hole potential $\mathbf V''$ (red) is screened to the bulk 
		value $\mathbf V$ far from it (pink), and is reduced to $\mathbf V_{\mathrm{TB}}$ in
		region 1 using orbital pruning (red/white pattern). $\mathbf V_{\mathrm{DFT-TB}}$
		connects the selected orbitals in region 1 to all orbitals in region 2.
		(b) An external unperturbed TB model, where the DFT-modeled hole from (a) is replaced by a self-energy $\boldsymbol\Sigma'$.
		(c) Comparison between transmission across a DFT two-electrode graphene device with
		a hole ($\mathcal T_{LR}$) and across a TB device with pristine scattering region and a
		self-energy $\boldsymbol\Sigma'$ substituting the DFT hole ($\mathcal T_{LR'}$). 
		The TB model is parameterized from $p_z$ orbitals of an unperturbed DFT calculation with SZ basis set. Four cases are shown where all the $p_z$ couplings from DFT, or only those up to $1^{\mathrm{st}}$, $2^{\mathrm{nd}}$ and $5^{\mathrm{th}}$ nearest neighbors, are retained in the TB parameterization.
		Insets shows schematically the procedure to calculate $\boldsymbol\Sigma'_{R'}$. Electrodes are highlighted in red and the opaque area shows the geometry which is replaced by $\boldsymbol\Sigma'_{R'}$.}
	\label{fig:dft2tbsketch}
\end{figure*}

%%%%%%%%%%%%%%%%%%%%%
\subsection{Self-energy for DFT-TB connections}
\def\DFT2TB{\mathrm{DFT-TB}}

We now turn to the more interesting situation where the perturbation and the external
unperturbed system hosting its self-energy $\boldsymbol\Sigma'$ are modelled using
\emph{different} basis sets, e.g. DFT and TB. We have already anticipated with the
example in \Figref{fig:grbands} that the electronic structure of a DFT system within a
particular energy window often can be reproduced by a small TB model obtained as a
projection of the DFT Hamiltonian. The drastically reduced number of orbitals in a TB
model makes it potentially very convenient for generating unperturbed external host
systems with large dimensions, normally inaccessible by DFT calculations. By generalizing the definition of the self-energy, $\boldsymbol\Sigma'$, it becomes possible to
incorporate DFT-precision perturbations in TB models. However determining a
self-energy for a DFT-TB connection is not as trivial as for the basic DFT-DFT connection
described in the previous section.

Consider now the system as shown in \Figref{fig:dft2tbsketch}a. Instead of having both
regions (1 and 2) using the same basis set, we change the basis set in region 1 to be a TB
parameterized basis set, $P\{\phi_{\alpha}\}=\{\bar{\phi}_{\alpha}\}$, as discussed in
\Figref{fig:grbands}. Here the approximation lies in $\mathbf V_{\DFT2TB}$. By the same
arguments outlined in the theory section, we create the subset of $\mathbf V_{\DFT2TB}$
such that the rows correspond to the full basis set $\{\phi_\alpha\}$ and the columns
correspond to the projection orbitals $P\{\phi_\alpha\}$. Using
$\mathbf V_{\DFT2TB}$ in \Eqref{eq:sigmap} results in a projection of the self-energy onto the parameterized orbitals.
Importantly, the fact that some elements of the original DFT coupling, $\mathbf V$, are now
missing in $\mathbf V_{\DFT2TB}=P\mathbf V$ inevitably leads to some scattering at the DFT-TB
boundary. 
%\MB{I do not understand this precisely:}
%One can minimize the impact on transport by parameterizing the TB model from DFT such that $\Sigma'_{\DFT2TB} \approx \Sigma_{\{\bar{\phi}_{\alpha}\}}$, as in this case DFT and TB bands match at best. As anticipated in \Figref{fig:grbands}c, overly simplified TB models would cause significant band-rescaling and misalignment and hence additional scattering.

%\N{Could we perhaps state that when $\Sigma_{\DFT2TB}'\approx \Sigma_{\{\phi_{\alpha}\}_{\Delta E}}$ this procedure is fine? Isn't this the requirement? However, for the NPG there is also band-misalignment, so then it does not become so obvious why it actually works?  }

\subsubsection{Example: graphene device with DFT-TB connection}

We have demonstrated that a TB model fully parametrized from DFT can reproduce the $\pi$
bands of graphene with only minimal deviations (\Figref{fig:grbands}c). As a result DFT
results can be reproduced, within accuracy, using the multi-scale DFT-TB approach.  Let us
consider again the graphene device with a hole, as shown in
\Figref{fig:dft2tbsketch}a. The only difference between DFT-DFT and DFT-TB is that here
$\mathbf V_{1,2}$, in \Eqref{eq:sigmap}, is chosen such that the columns of
$\mathbf V_{1,2}$ project onto the parameterized $p_z$ orbitals. This yields a
self-energy only existing on the $p_z$ orbitals in the TB model. However, one may
additionally play with the number of neighbors each atom connects with,
e.g. 1, 2, 3 or all neighbors (with
``all'' defined by the DFT basis).

In \Figref{fig:dft2tbsketch}c the transmission for the hole system is shown at four
levels of precision compared to the full DFT calculation. 
We find that if the TB parametrization contains only DFT couplings among nearest neighbors then the transmission spectrum undergoes significant scattering $\sim$0.1 eV away from the Dirac point. When the range of interaction is extended to 2, 3 or 4 neighbors the transmission spectrum becomes comparable to DFT in a wider energy range, except for a shift at positive energies. As soon as couplings among 5 nearest neighbors are included we obtain an almost perfect agreement with DFT.

%%%%%%%%%%%%%%%%%%%%%
\subsection{Self-energy of isolated perturbations}

In all the examples considered so far \emph{periodic} boundary conditions was employed in the 
DFT calculations from which the connecting self-energy
$\boldsymbol\Sigma'_{\DFT2TB}$ was constructed. This means that the perturbation, e.g. the hole, is
periodically repeated along the transverse direction $y$. However, if the potential is screened to the bulk value one may replace the surrounding periodic images by other environments with the same potential and Hamiltonian. In this way one may effectively change the boundary conditions.

%\MB{I think the data-technical point of "removing PBC" from siesta data in sisl should not be mixed with the physics (screening). The central assumption is that the potential in a "frame" away from the perturbation is bulk and thus can be connected to other systems where this again has converged to bulk.}

The origin of this versatility lies in the fact that the two regions in which the
perturbed DFT model is divided can have arbitrary size, shape and periodicity. For
example, with reference to \Figref{fig:dft2tbsketch}, one could choose to define the
pristine DFT region 1 as the outermost frame-shaped area of the cell surrounding the
perturbation, while treating all the orbitals enclosed by it as region 2. 
The same formalism discussed above can be readily applied to
this case, with the only difference that periodic boundary conditions are removed from
the DFT Hamiltonian \emph{before} constructing the self-energy $\boldsymbol\Sigma'_{\DFT2TB}$.

\begin{figure}[t]
	\includegraphics[width=0.9\columnwidth]{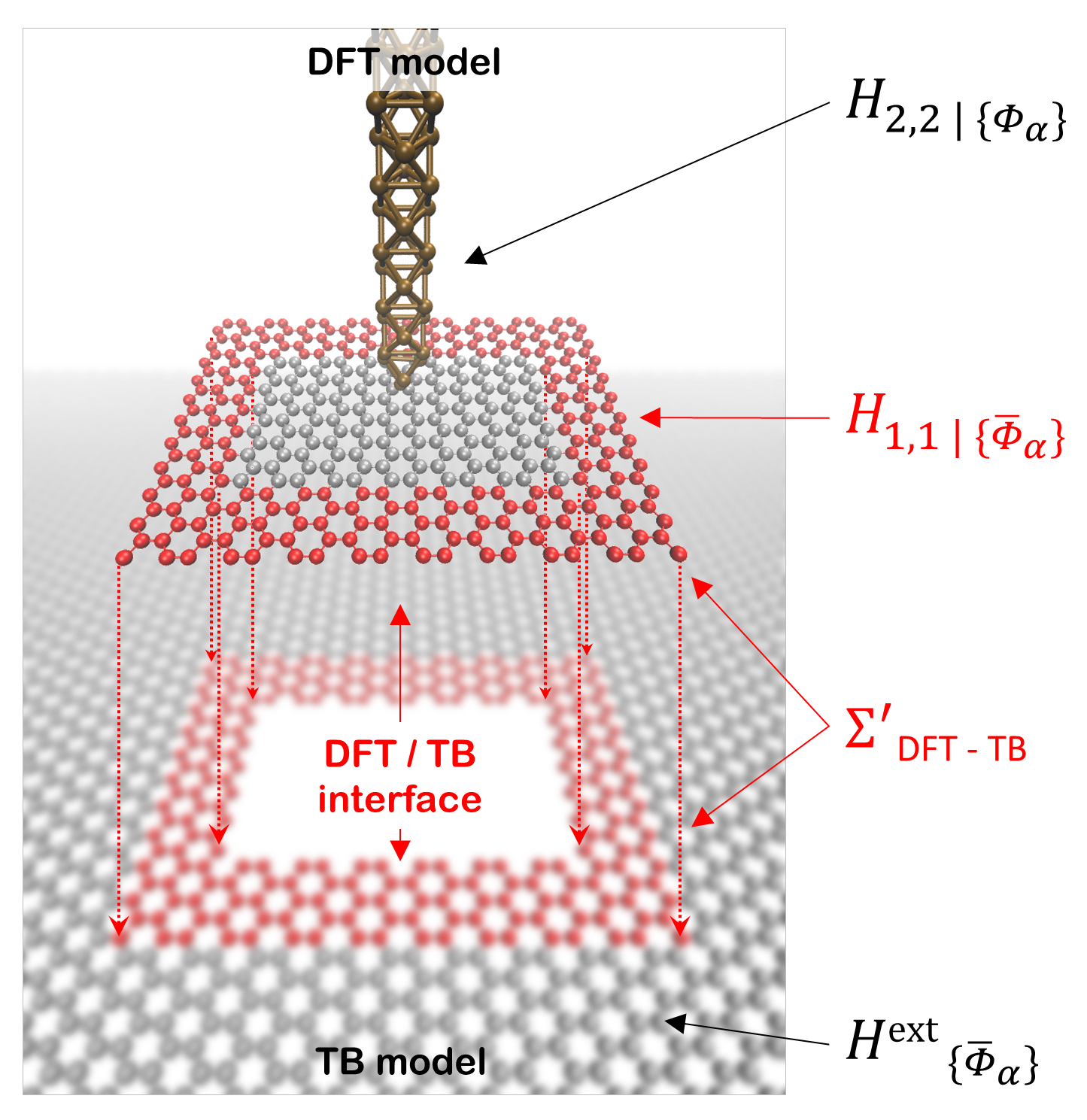}
	\caption{Illustration of non-periodic calculation of $\boldsymbol\Sigma'_{\DFT2TB}$
		enabling studies of far-field effects.
		Top geometry is the DFT simulation cell, while the bottom is the extended
		parameterized TB model. The DFT region is split into two regions, $1$ and $2$ as
		indicated via $\mathbf H_{i,i}$, with the first region only retaining the
		parameterized orbitals.
		The projected self-energy is calculated on the marked atoms in the DFT system and
		transferred as an electrode into the TB model, thus retaining DFT accuracy of the
		electronic structure between the tip and graphene. 
		\label{fig:framepristine}
	}
\end{figure}

An example is illustrated in \Figref{fig:framepristine}. The top geometry is the DFT region which is used in calculating $\boldsymbol\Sigma'_{\DFT2TB}$. The bottom
geometry is the TB parameterization and the red atoms indicate the overlay region where
the self-energy is transferred from region 1 (DFT) to region 2 (TB).
We point out that this particular approach provides the basis for ``modular'' multi-scale
simulations, where multiple DFT-precision perturbations, \emph{modules}, are incorporated
into the same large TB device. We will present concrete applications of this in the second
part of the work.

%%%%%%%%%%%%%%%%%%%%%
\subsection{Challenges and implementation}

Despite reflecting a conceptually simple theory, the protocol to embed
DFT-precision perturbations locally into an unperturbed TB region parametrized from DFT
implies a number of critical issues when it comes to its practical implementation.
Let us briefly go through the procedure to include non-periodic DFT perturbations
into large TB models step-by-step:
\begin{enumerate}
	\item Generate a DFT model of the perturbation. This can be represented by either
	a bulk (periodic) or device (semi-infinite electrodes) setup and must be large enough to ensure the potential turns constant in its outermost
	areas;
	
	\item Define two regions in the DFT system, one containing all orbitals involved
	in the perturbation and the other involving the projection orbitals. The latter needs to
	be as far as possible from the perturbation so as to ensure $\mathbf V_{\DFT2TB} =
	\mathrm{constant}$;
	
	\item Represent the DFT Hamiltonian and
	overlap in real-space (without periodic boundary conditions);
	
	\item Compute the self-energy $\boldsymbol \Sigma'$ using equation
	\Eqref{eq:sigmap};
	
	\item Generate a DFT model of the same system, but without the perturbation;
	
	\item Construct a large TB model using parameters from the projection of the unperturbed
	DFT calculation.
	
	\item Incorporate the self-energy $\boldsymbol \Sigma'$ locally into the TB model using Eq.~\eqref{eq:green}.
	
	%\item Use it \MB{What is "it"??} \GC{I joined this and the previous points} as electrode in a transport calculation using the NEGF equations Eqs.~\eqref{eq:t} and \eqref{eq:green}.
	
\end{enumerate}

To begin with, most of the described steps require flexible manipulation of the DFT
Hamiltonian and overlap matrices. Extraction of selected on- and off-diagonal elements
exclusively associated to a subset of orbitals is crucial to compute
$\mathbf G_{2,2}$, extract $\mathbf V_{\DFT2TB}$ or to create TB models parametrized
from DFT.
Secondly, to ensure the correct couplings at the DFT-TB interface when incorporating
$\boldsymbol\Sigma'$ into the TB model, it is important that region/atoms selected
to host $\boldsymbol\Sigma'$ in the large TB geometry have a one-to-one correspondence
with those in region 1 of the DFT geometry. 
%\MB{Think if we use a general $P$.. then I guess this is more that the atoms are the same?}
User-friendly tools to access and
compare lattice coordinates of the various involved geometries are therefore highly
desirable.

Furthermore, the self-energy, $\boldsymbol\Sigma'$, needs to be stored into external files
with general and compact format, readily accessible for usage in NEGF calculations. Being
able to flexibly input/output self-energies into external host models is especially
crucial to carry out multi-scale calculations where several DFT-precision
regions are accounted for in the same large TB device.

Overall, the computation of $\boldsymbol\Sigma'$ through \Eqref{eq:sigmap} and NEGF
transmissions needs to be optimized both time and memory-wise, so as to efficiently handle
dense DFT matrices and large sparse TB ones, as well as calculation of the self-energy for
a fine $k$ grid and several energy values.

\paragraph*{Computational tools}
We tackle all the issues described in the previous section with open-source tools
\tbtrans\ and \sisl.\cite{Papior2017,Papior2017a} \tbtrans\ is distributed with the \siesta/\tsiesta\ software package.\cite{Soler2002,Papior2017}
\tsiesta\ enables high-performance DFT+NEGF self-consistent calculations of large,
multi-terminal systems under various electrostatic conditions (e.g. gating.\cite{Papior2015}) \tbtrans\ is a post-processing NEGF code which provides a flexible
interface to DFT or TB Hamiltonians and enables large-scale calculations of spectral
physical quantities, interpolated $I$-$V$ curves and/or orbital/bond-currents for systems
easily exceeding millions of orbitals on few-core machines. \sisl\ is a Python package
used to create and/or manipulate DFT and TB models for arbitrary geometries, with any
number of orbitals and any periodicity.\cite{Calogero2018}
The device Green's function in \tbtrans\ is generally implemented as:
\begin{equation}
\label{eq:tbt-G}
\mathbf G_D = \left[ \mathbf S\,(E + i\eta) - \mathbf H - \sum_i\boldsymbol\Sigma_i - \delta\boldsymbol\Sigma \right]^{-1}
\end{equation}
where $\mathbf S$ and $\mathbf H$ are DFT or TB-modeled overlap and Hamiltonian, $\boldsymbol\Sigma_i$ are the electrodes self-energies and $\delta\boldsymbol\Sigma$ is a user-defined, optional, perturbative term. 
%All these terms depend on energy and $\mathbf k$.  We typically set up $\delta\boldsymbol\Sigma$ as an absorbing potential (CAP) absorbing electrons at side walls of the large unperturbed TB model \cite{Calogero2018, Xie2014}.
The flexibility of using \sisl\ and \tbtrans\ is appreciated since no code extension of
\tbtrans\ is required as everything being done is added using $\delta\boldsymbol\Sigma$
and $\boldsymbol\Sigma_i$ terms. Implementation of these two terms in \Eqref{eq:tbt-G} is
done \emph{only} in Python and is thus much simpler than a full fortran
implementation. Secondly, both \sisl\ and \tbtrans\ are scaling easily to millions of
orbitals in the TB method.

% Resources
\change{All of the calculations described in this article have been carried out using a single high-performance computing node with 20 cores (Core Intel Xeon E5-2660v3, 2.6GHz, 128GB RAM), parallelizing \tbtrans\ and either MPI or OMP parallelization. The transmission spectra in \Figref{fig:se} and \Figref{fig:dft2tbsketch} were calculated in \tbtrans\ using 100 energy points, 201 k-points in the periodic direction and a total of 540 atoms. The computationally most expensive calculation presented so far, i.e.\,, that involving a DFT-DFT connection, required 3 hours and 50 minutes to compute/store the connecting self-energy $\boldsymbol\Sigma'_{\DFT2TB}$, and 10 minutes (53MB RAM, 113kB disk) to calculate transmission with \tbtrans.} 

%%%%%%%%%%%%%%%%%%
\section{Applications}
In the second part of this work we show how the multi-scale approach can be used to study the far-field behavior of electrons injected from atomic-scale contacts into large-scale graphene-based devices.

Being able to interpret and predict the behavior of electrons over large-scale devices is of importance for graphene-based electronics, especially in the growing field of electron optics. Real-space visualization of charge (or spin) transport can be achieved experimentally using various quantum imaging techniques, including probe microscopy,\cite{Hiraoka2017, Kolmer2017} superconducting interferometry\cite{Allen2015} and magnetometry with diamond-NV centers.\cite{Casola2018} It has been shown that defects and contacts with tips at the atomic scale yield strong spatial variations of current flow in graphene devices.\cite{Bhandari2016, Tetienne2017} 

Modeling realistic atomic-scale contacts to inject currents in graphene devices, while simultaneously accessing current flow far away from them, is a challenge for state-of-the-art atomistic transport calculations. Large graphene devices can be accessed using the TB approximation. Despite allowing for good scalability, in this context the complex chemical nature of defects or tip contacts is often subject to drastic simplifications. For instance a STM probe is usually modeled as a constant, on-site, level broadening ($i\Gamma$) self-energy term in the Green's function,\cite{Aprojanz2018, Settnes2014} a localized effective force or electrostatic field,\cite{Eder2013, Mark2017} or even as a narrow semi-infinite in-plane electrode.\cite{Calogero2018}
On the other hand, DFT-based models have often demonstrated to successfully corroborate probe microscopy measurements on graphene, even when involving complex tip functionalizations or inelastic effects.\cite{Jelinek2017, Frederiksen2011, Palsgaard2015,Heijden2016} This is due to an accurate description of tip structure and orbital symmetries, as well as charge and potential variations in the sampled regions.

The essential need for both scalability and accuracy in this  problem calls for the multi-scale approach described above.
In the following we will provide a bird's eye view on carrier injection from point sources into pristine, defected and nano-porous graphene (NPG). Furthermore, In order to further emphasize the versatility of the method, we will show an example where multiple DFT-precision regions from various DFT calculations are included in the same large-scale TB transport calculation.

%%%%%%%%%%%%%%%%%%%%%
\subsection{Far-field currents in graphene}
\begin{figure}[t]
	\includegraphics[width=1.0\columnwidth]{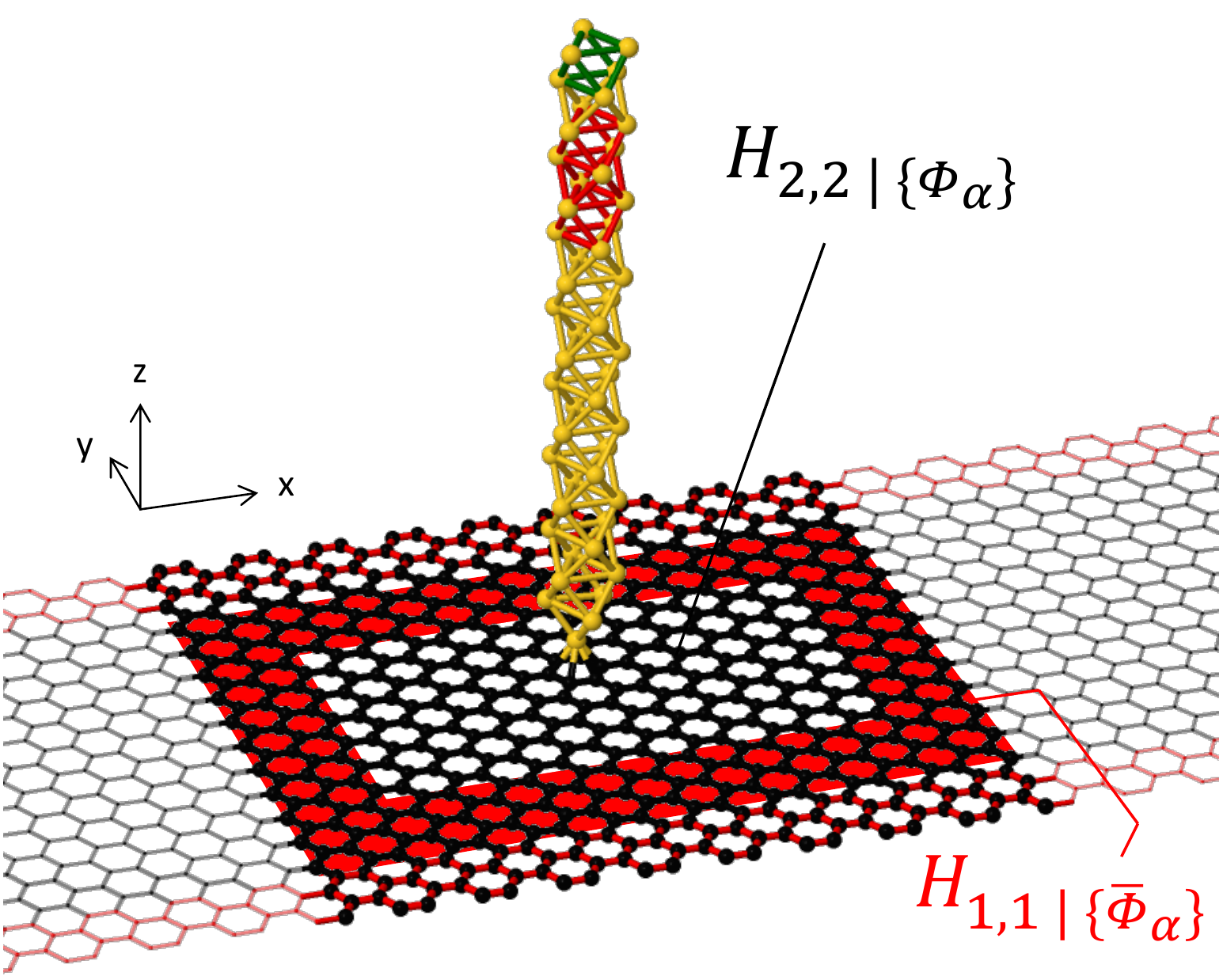}
	\caption{
		DFT model of our three-probe graphene-based device in contact to a model STM tip. The cell is periodic along $x$. Atoms defining the two graphene electrodes (semi-infinite along $y$) and the tip electrode (semi-infinite along $z$) are indicated with red bonds. The red underlying area indicates region 1 only retaining the parameterized orbitals, while the rest of the device represents region 2, whose degrees of freedom will effectively be replaced with the self-energy $\boldsymbol\Sigma_{\DFT2TB}$. Green bonds in the tip indicate ``buffer'' atoms.\cite{Papior2017}}
	\label{fig:setuppristine}
\end{figure}

\begin{figure}[t]
	\includegraphics[width=1.0\columnwidth]{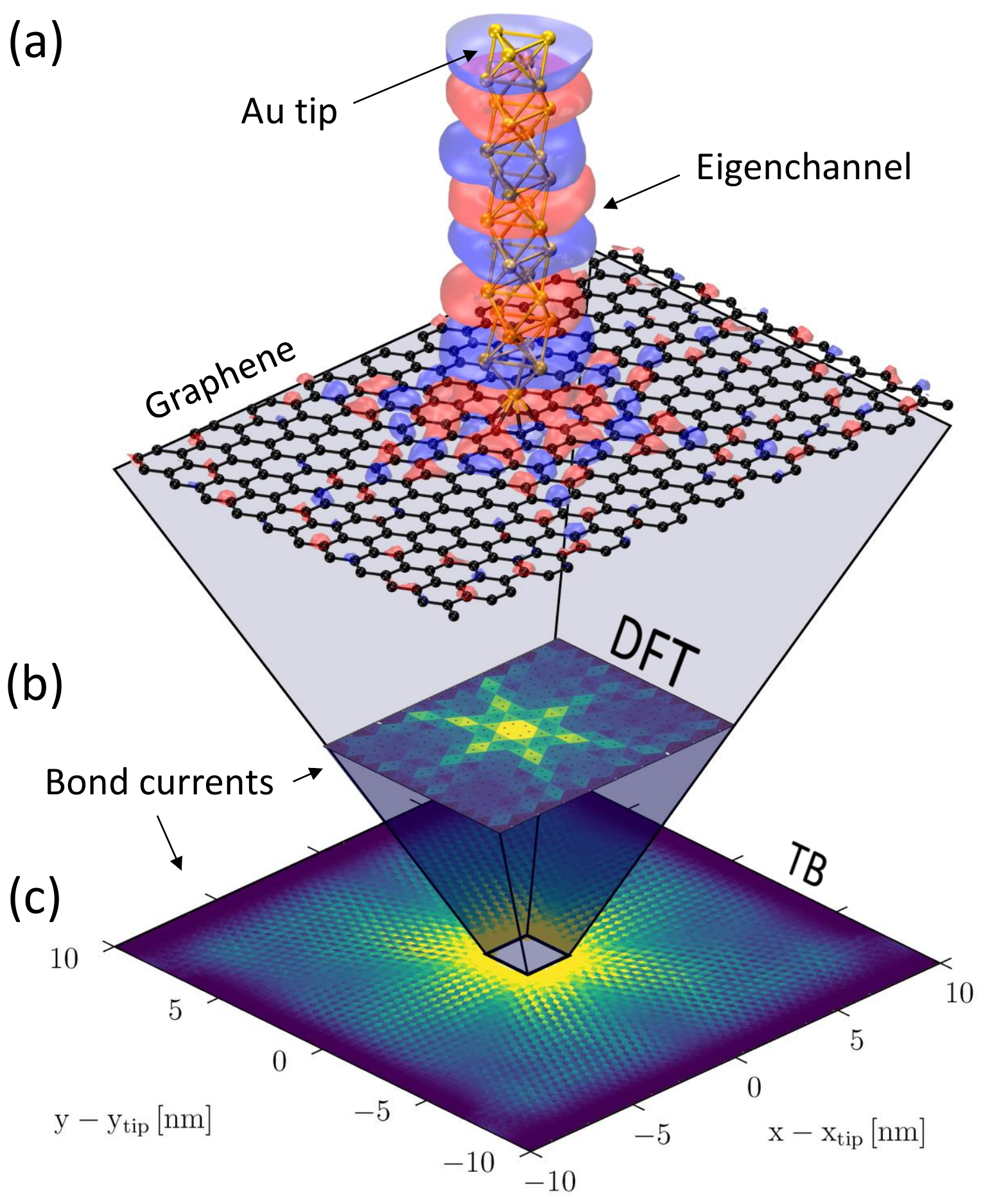}
	\caption{
		(a) DFT geometry of a model gold tip in contact to pristine graphene. Real and imaginary part of the anisotropic transmission eigenchannel, existing at $E-E_{\rm F}=0.8\,\ev$ between the tip and two graphene electrodes along $y$, are shown in red and blue. 
		(b) Bond-currents injected by the tip at $E-E_{\rm F}=0.8\,\ev$, calculated using a real-space self-energy in the outermost DFT region. 
		(c) Far-field bond-currents injected by the DFT-precision tip into a large-scale TB model of graphene parameterized from DFT.}
	\label{fig:tippristine}
\end{figure}

Here we use the multi-scale approach to predict the far-field behavior of electrons injected by a DFT-precision gold tip into a large-scale TB model of pristine graphene. 
Using wave-packet dynamics simulations to inject currents from an effective electrostatic field model of STM tip, Mark \etal reported anisotropic electron current in graphene.\cite{Mark2012}
The authors predicted that a well-defined six-fold asymmetry current pattern occurs when the wave packet is injected in the middle of a hexagon in the graphene lattice.\cite{Mark2012} Below we confirm this anisotropic signature by looking at the far-field currents flowing up to 10 nm away from the injection. The system is modeled by TB and is coupled to a small DFT-precision injection region through a self-energy $\boldsymbol\Sigma_{\DFT2TB}$. We will limit our focus to a single representative energy value where this six-fold anisotropy is clearly visible. 
%A full study of transport over the full energy spectrum is beyond the scope of this article.

%\MB{For the next we should extend Fig. 5 since it becomes too difficult to understand the setup. We need also to explain that we remove the PBC by summing over all transverse k-points -- we need formulas. The initial DFT relaxation is standard with PBC .. how many k-points? }\GC{I will make a figure with the setup. Explicitly mention that relaxation is performed with PBC and ?? kpoints. And also insert below the kavg formula from old version of the manuscript!!!}
We start by setting up a three-electrode DFT model device (see \Figref{fig:setuppristine}) consisting of a semi-infinite gold tip placed $\approx2.0\,\angstrom$ above the center of a graphene hexagon and two semi-infinite graphene electrodes along $y$. The tip structure is chosen so as to ensure a flat local density of states on the tip apex in a wide energy interval, $[-0.75 , \, 1] \,\ev$. We optimize the tip apex atom and the nearest $\sim20$ C atoms in \siesta\ until forces are less than $0.01\,\ev/\angstrom$, using periodic boundary conditions along $x$ and $y$, a $3\times3$ $k$-point Monkhorst-Pack grid, the GGA-PBE exchange-correlation functional,\cite{Perdew1996} a SZP basis set and an energy cutoff of 300 Ry. \change{The tip is in chemical contact to graphene, therefore we assume van der Waals interactions to be negligible and do not include them in the calculations.}
We then calculate bond-currents flowing across the three-electrodes device using the tip as a source, using $30$ $k$-points to sample the $x$ direction in \tbtrans.
In order to avoid artificial crosstalk with periodic images of the tip when computing the injected bond-currents, we substitute the self-energies for the two graphene electrodes with a new self-energy term $\boldsymbol\Sigma_{\rm avg}$ in \Eqref{eq:green},
\begin{equation}\label{eq:SEinside}
\boldsymbol\Sigma_{\rm avg} = \langle \mathbf S \rangle_k (E+i\eta) - \langle \mathbf H \rangle_k - \langle \mathbf G \rangle_k^{-1} \,,
\end{equation}
where $\langle\cdot\rangle_k$ represents an average over all $k$-points along $x$, i.e.\ the quantities in the principal unit cell in real-space.\cite{Papior2019}
In effect, this is equivalent to setting up a new drain electrode in the border region of the cell.
%Further details on this approach will be reported elsewhere \GC{Maybe cite Nick as in preparation? Also, should I put the image with bond-currents with and without elimination of PBC? Or should we leave it for the new paper?}\MB{Leave it for the new paper .. things are done differently with Bloch-repetion there anyways.} \GC{Insert kavg formula from old version of manuscript}. 

The results for $E-E_{\rm F}=0.8\,\ev$ are summarized in \Figref{fig:tippristine}. 
We find that the total transmission from tip to $\boldsymbol\Sigma_{\rm avg}$ is $\mathcal T=0.686$. This is determined mainly by three transmission eigenchannels\cite{Paulsson2007}. The two with largest contribution (37\% and 34\% of the total transmission) are mostly delocalized over the whole graphene structure, whereas the third one (contributing with 29\%) exhibits a preferential propagation along the six armchair lattice directions departing from the probed graphene hexagon (\Figref{fig:tippristine}a). We find that this six-fold anisotropy dominates the bond-current pattern injected by the tip at this energy (\Figref{fig:tippristine}b).

Then, using the approach discussed in \Figref{fig:framepristine}, we insert the DFT-precision injection region inside a larger TB model parameterized from DFT, using a complex absorbing potential\cite{Calogero2018, Xie2014} (CAP) to absorb currents at the cell boundaries.
As shown in \Figref{fig:tippristine}c we find that the six-fold anisotropic propagation can still be observed far away from the source, although collimation is rapidly lost. 
As a check we estimate transmission outgoing from the tip by summing over positive bond-currents which cross a circle of radius $R$ centered at the tip $(x,y)$ coordinates. The result, $\mathcal T=0.672$, is only slightly lower than the value obtained by summing bond-currents in the smaller all-DFT system, $\mathcal T=0.681$, regardless the choice of radius $R$.

\begin{figure}[t]
	\includegraphics[width=1.0\columnwidth]{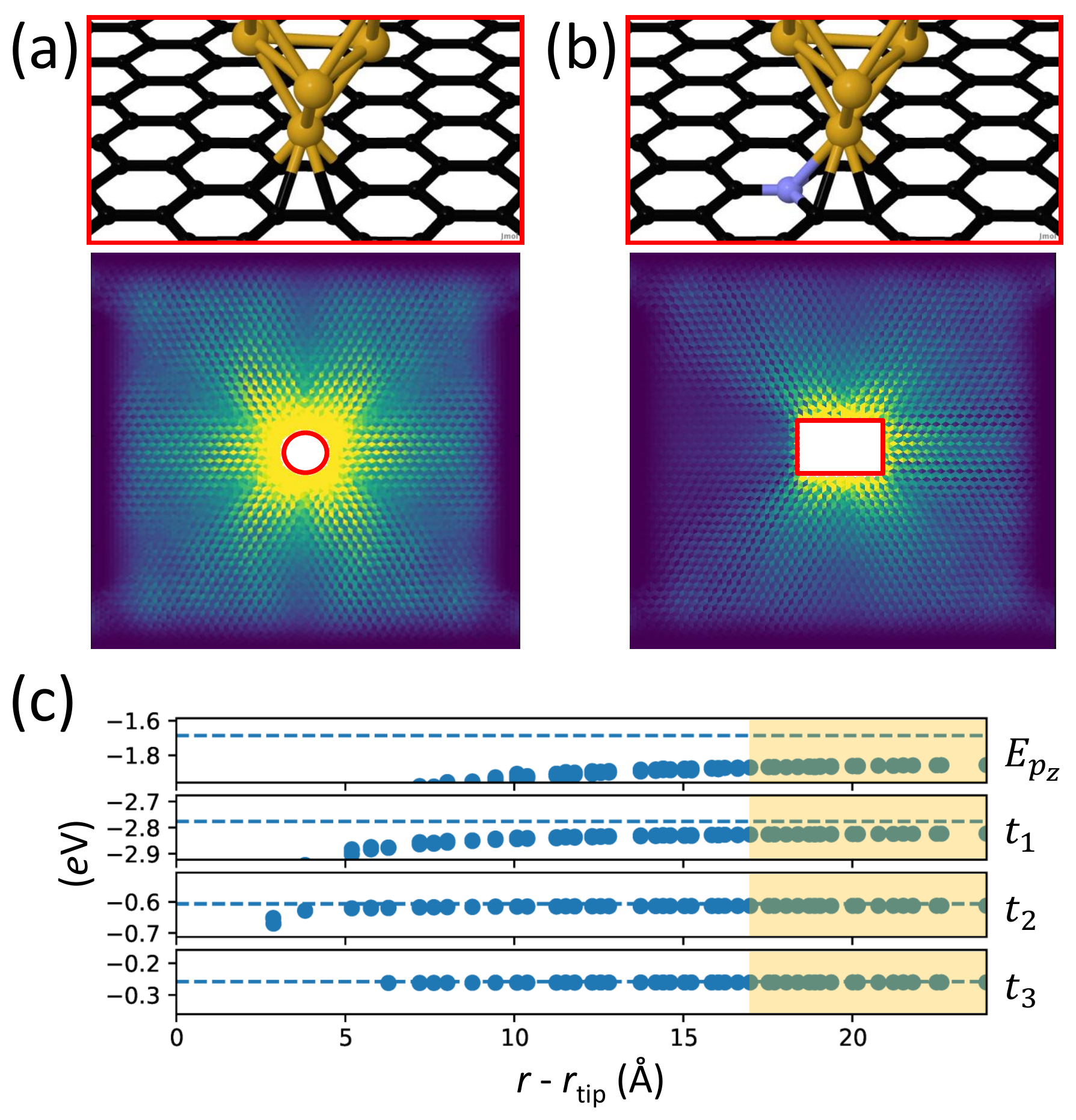}
	\caption{
		(a) Far-field bond-currents injected at $E-E_{\rm F}=0.8\,\ev$ by a model STM tip into a large-scale TB+DFT model of graphene, obtained using a circularly shaped outermost DFT region to compute $\boldsymbol\Sigma_{\DFT2TB}$. The result is the same as that obtained with a rectangular shape.
		(b) Far-field bond-currents injected by the tip at the same energy into a TB+DFT model of N-substitutional dopant in graphene, where the N dopant sits on one of the sites contacted by the tip.
		(c)  On-site potential and couplings among C-$p_z$ orbitals in the DFT model of STM contacted N-doped graphene as a function of $xy$-distance from the tip. Beyond $\approx 17\,\angstrom$ the couplings are approximately constant, and different from the average ones expected from a DFT calculation of non-contacted pristine graphene (dashed). The $y$ axis is chosen to emphasize this difference at a large distance (hence some of the data at small $r-r_{\rm tip}$ are out of the $y$-axis range).
	}
	\label{fig:N}
\end{figure}

Importantly, we find that the result does not change if the shape of the DFT region in which $\boldsymbol\Sigma_{\DFT2TB}$ is computed is modified from a rectangular to a circular one (\Figref{fig:N}a).

\paragraph*{Effects of single N dopant on far-field currents}
One major advantage of the multi-scale approach is that complex situations, e.g. where different perturbations simultaneously affect the same region, can be described with chemical accuracy. An illustrative example is depicted in \Figref{fig:N}b, where the multi-scale method has been used to study how a N substitutional dopant in proximity of the Au tip contact affects the far-field currents. 
In order to unravel possible effects of spin polarization we carry out this calculations for its two possible spin configurations.
We find that in both cases, due to the local doping induced by the N atom, electrons injected by the tip are prevented from propagating towards the N atom (\Figref{fig:N}c).
%\MB{Should we actually plot the eigenchannels close to the tip with and without N?} \GC{That is a very good idea}

\paragraph*{TB parameterization} We point out that, compared to the simple case discussed in \Figref{fig:dft2tbsketch}, the presence of a metal contact in these systems alters the graphene work-function, and hence its potential and couplings far from the tip. As a result, parameterizing the TB model directly from a DFT calculation of pristine graphene would induce some degree of coupling mismatch at the interface between the two models. The solution we adopt here to ensure constant coupling $\mathbf V_{\DFT2TB}$ is to parameterize the TB model from the pristine-like DFT elements associated to the atoms far away from the tip, namely those further than $\approx 17\,\angstrom$ from it (\Figref{fig:N}c).

%%%%%%%%%%%%%%%%%%%%%
\subsection{Far-field currents in nanoporous graphene}
\begin{figure}[t]
	\includegraphics[width=1.0\columnwidth]{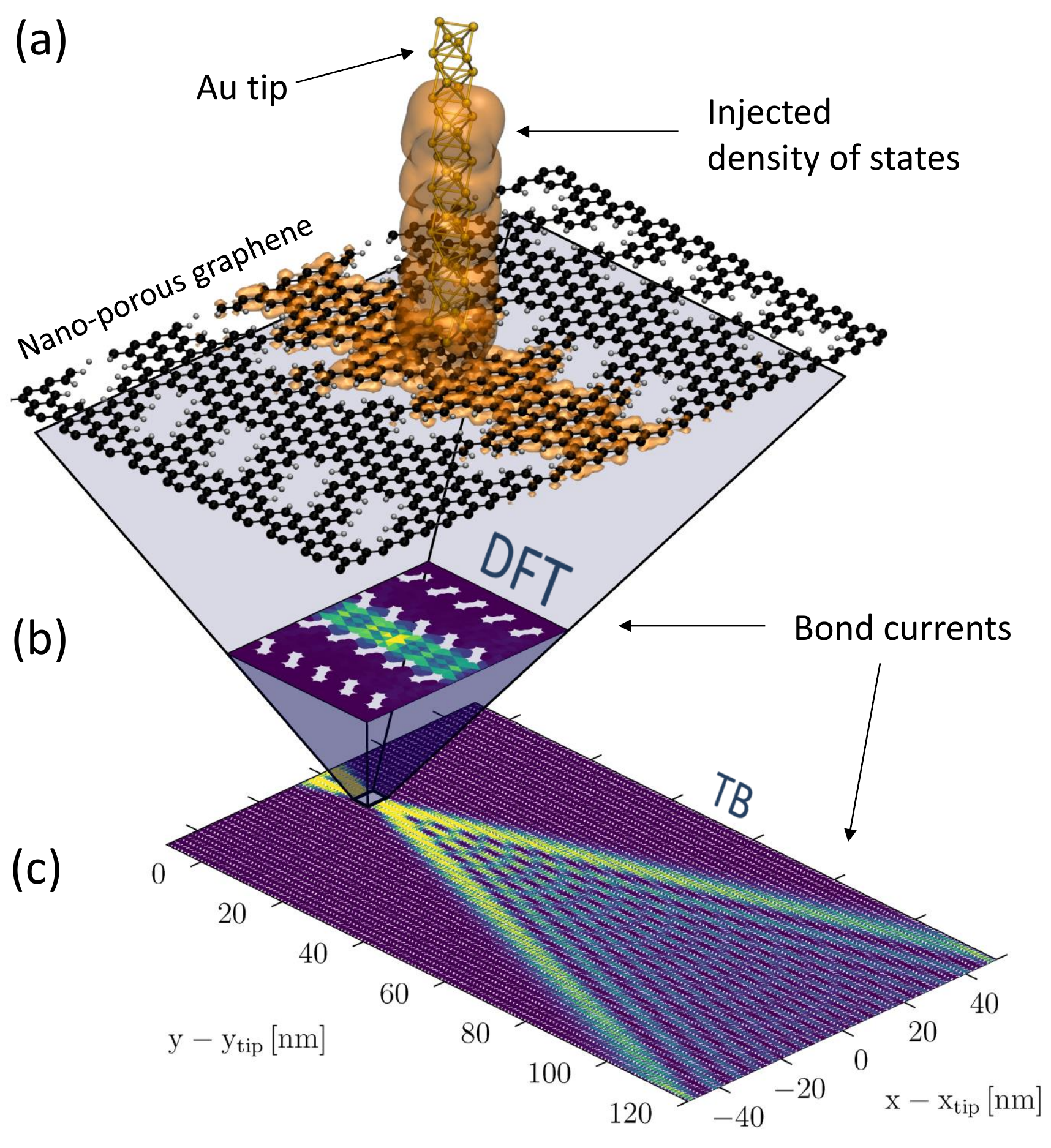}
	\caption{
		(a) DFT geometry of a model STM tip in contact to NPG. The system has periodic boundary conditions along $x$ and two semi-infinite electrodes along $y$. The density of states injected from the tip electrode and reaching the two NPG electrodes is shown in orange. 
		(b) Bond-currents injected by the tip into one of the ribbons making up the NPG at $E-E_{\rm F}=0.2\,\ev$.
		(c) Far-field bond-currents injected by the DFT-precision tip into a large-scale TB model of NPG parameterized from DFT.
		Figure adapted with permission from Ref.~[\citenum{talbotNPG} ], Copyright 2019 American Chemical Society.
		\label{fig:npg}}
\end{figure}
The principles behind the multi-scale method are general and not limited to the basic graphene systems presented above.
For instance, let us now consider nanoporous graphene (NPG), which is inherently semiconducting and anisotropic.\cite{Pedersen2008} In particular we consider the NPG structure shown in \Figref{fig:npg}a, which has recently been fabricated through bottom-up synthesis with unprecedented sample sizes and quality.\cite{Moreno2018} In this system an electron wave is forced to channel into coupled 1D parallel pathways, in analogy to light waves propagating through an array of optical wave-guides. In a recent work\cite{talbotNPG} we have exploited the capabilities of the multi-scale approach to tackle two essential questions in this context, namely whether these transmission channels interfere with each other and how the wave profile can be tuned and controlled over long propagation distances. We have demonstrated that an electron wave injected into NPG from a model STM tip spreads over the nanomesh according to the same equations which govern the so-called Talbot optical interference effect (\Figref{fig:npg}b-c). Further information can be found in Ref.~[\citenum{talbotNPG}].

Application of the multi-scale method in this context was vital, as the typical distances accessible by DFT are not long enough to capture the characteristic interference fringes (see \Figref{fig:npg}b).
Importantly, contrary to the graphene systems considered in the previous section, in this study we have ensured constant coupling $\mathbf V_{\DFT2TB}$ by introducing doping via an electrostatic gate $15\,\angstrom$ below the NPG plane, such that  $\pm10^{13}\,e^-/\cm^2$ carriers are induced in the NPG.\cite{Papior2015} This ensures that the Fermi level is pinned throughout the whole DFT+TB device, effectively avoiding any possible artificial mismatch between the models involved in the calculation of $\boldsymbol\Sigma_{\DFT2TB}$.

%%%%%%%%%%%%%%%%%%%%%
\subsection{Large-scale TB with N>1 DFT-precision regions}
\begin{figure*}[t]
	\includegraphics[width=1.0\textwidth]{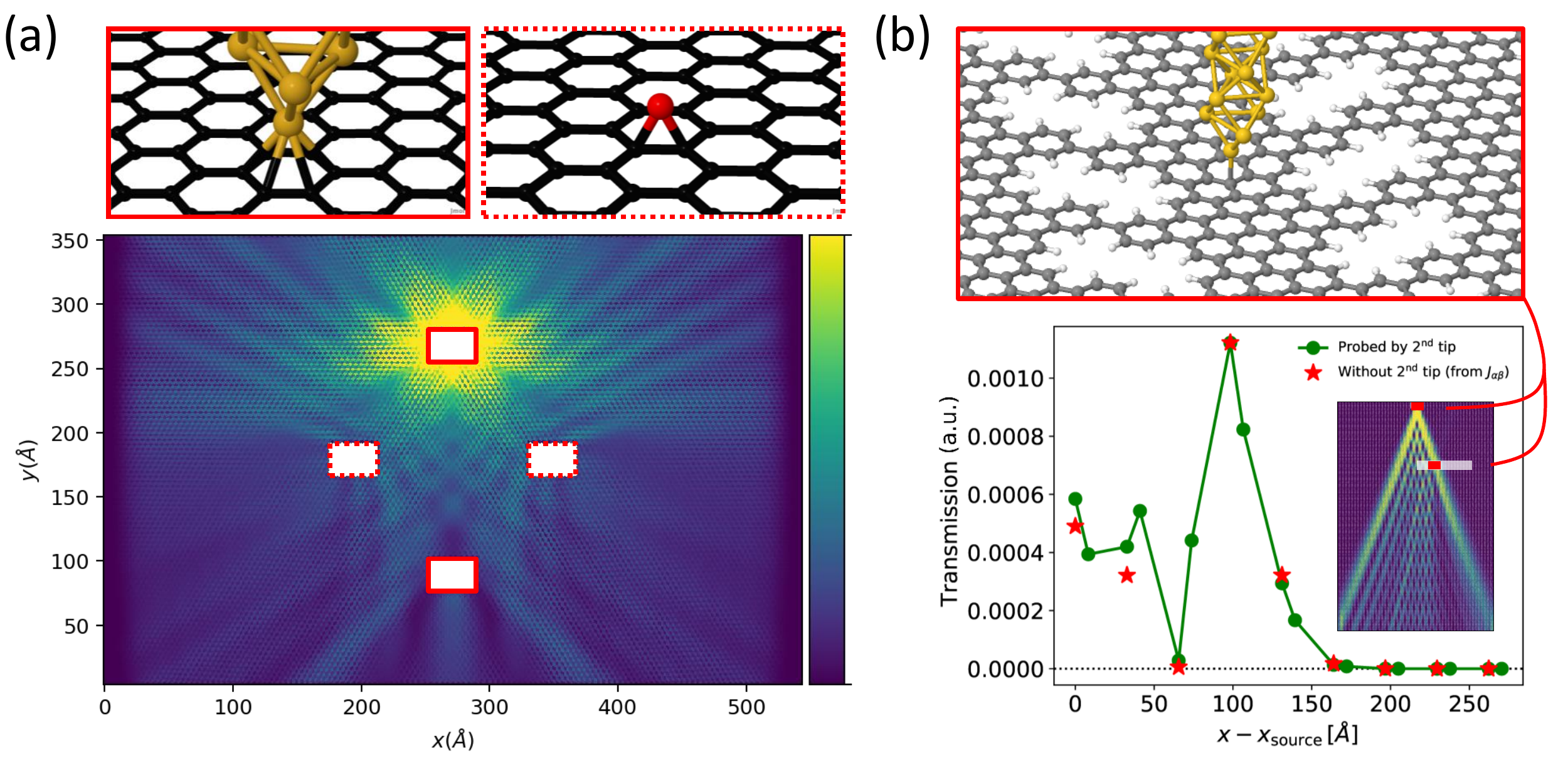}
	\caption{
		(a) Far-field currents from a multi-scale calculation of a TB graphene device parameterized from a DFT calculation of gated graphene, with $N=4$ DFT-precision regions, i.e. two STM tip contacts (solid red rectangles) and two epoxy defects (dotted red rectangles). We use two semi-infinite electrodes along $y$ and CAP along $x$. Bond currents are injected at $E-E_{\rm F}=0.8\,\ev$ from the upper tip. Colors are scaled to enhance contrast.
		(b) Transmission at $E-E_{\mathrm F}=0.2\,\ev$ (green) measured by a $2^{\mathrm{nd}}$ DFT-precision tip scanning over a gated NPG
		along the white line shown in inset, in comparison to bond-currents flowing in absence of the $2^{\mathrm{nd}}$ tip (red). 
		These are obtained on a ``per ribbon'' basis by summing all bond-currents passing through the white line, without distinction between ribbon and bridge sites, and then scaling by a factor $1/16$.
		The inset shows currents scattering off the $2^{\mathrm{nd}}$ tip in one of the scanned positions. Figure adapted with permission from Ref.~[\citenum{talbotNPG}], Copyright 2019 American Chemical Society.
		\label{fig:many}	
	}
\end{figure*}

Another advantage of the multi-scale method is that one can accommodate \emph{any} number of perturbed regions in the TB model, by simply including a corresponding number of self-energy terms in \Eqref{eq:green}. This is exemplified in \Figref{fig:many}, where we show results from multi-scale TB+DFT calculations where multiple DFT-precision regions have been set up within a single large-scale TB model.
\Figref{fig:many}a illustrates far-field currents from a multi-scale transport calculation of a $55\times35 \,\nm^2$ TB graphene device with seven self-energy terms. Two of these model regular graphene electrodes semi-infinite along $y$. The third one simulates a CAP in the outermost $5\,\nm$ of the cell along $x$. The remaining $N=4$ represent DFT-precision regions, namely two Au tips in contact to graphene ($\approx 2\,\angstrom$ above an hexagon) and two epoxy groups, which preserve the $sp^2$ nature of graphene\cite{Sljivancanin2013} and are of particular interest for e.g.\ engineering the thermal conductivity\cite{Zhao2015} or catalytic activity\cite{Sinthika2015} of carbon/graphene-based systems.
Both DFT systems are described using a SZP basis set and have been optimized with force threshold $0.01\,\ev/\angstrom$. Self-energies are computed once for each of the two systems, while a bottom gate induces $-10^{13}\,e^-/\cm^2$ into graphene. Bond-currents injected at $E-E_{\rm F}=0.8\,\ev$ by the upper Au tip into the large TB model clearly show how the six-fold anisotropic electron wave scatters off the defects and the second tip. 
We point out that, once the corresponding self-energies have been computed and stored, the DFT-precision regions can be moved around the large-scale TB model very efficiently. In the Supplementary Material we show how the bond-currents landscape looks like for different positions of an epoxy defect relative to two STM tip contacts.

%\MB{Maybe leave the (b) part out since it is already in the NanoLett and just write it in words in the beginning of the section as an example.}
Another interesting situation, already discussed in Ref.~[\citenum{talbotNPG}] is depicted in \Figref{fig:many}b, with regard to the Talbot interference pattern discussed in the previous section. By inserting a second STM tip into the large-scale TB model it is possible\cite{talbotNPG} to map out the interference pattern observed in the bond-currents by performing a dual-probe experiment where a second STM tip is scanned over the NPG structure.

\change{The computationally most expensive calculation presented in this section, i.e., the one illustrated in \Figref{fig:npg}, required (for 150 energy points, $\Gamma$ only) ~110 minutes to compute/store the connecting self-energy $\boldsymbol\Sigma'_{\DFT2TB}$, and ~4 hours (55GB RAM, 15GB disk) to calculate bond-currents with \tbtrans.}

% Fig.6; 36 sec python; 9 sec, 450MB RAM, 8MB disk TBT
% Fig.7b; 58 sec python; 15 sec, 790MB RAM, 5MB disk TBT
% Fig.8; 110 min (75Epts) --> 88 sec python; 55GB RAM; 15GB disk TBT

%%%%%%%%%%%%%%%%%%%%%
\section{Conclusions}
In conclusion, we have presented a multi-scale method which enables calculations of local currents in devices larger than 100 $\nm^2$, by linking a perturbed region described by DFT to an unperturbed large-scale region described by an effective TB model parametrized from DFT. 
We have introduced the theory behind the method using basic concepts in the Green's function framework, provided didactic examples and pointed out the main difficulties connected with its implementation.
By applying the method to study realistic current injection by STM probes into pristine, defected and nanoporous graphene devices we have highlighted versatility, efficiency and generality of the method.
Similar to hybrid QM/MM techniques, combining the advantages of
DFT and TB methodologies using this scheme provides an adequate
framework for embedding regions where accuracy is necessary into regions where size matters more.
\change{The largest system size accessible using this scheme obviously depends on the available computational resources and, most importantly, on the parametrization of the TB model, which varies with the material under study. We have illustrated the method in the simple case of carbon-based 2D materials, where the mapping from DFT to TB is a simple projection. Further studies may unveil whether other projectors are better or more generalizable. Overall, the most critical point is to define the TB to be computationally manageable, while coupling the pristine TB and DFT regions such that the interface scattering is much smaller than the scattering mechanisms under study.}
%We have illustrated the method in the simple case where the mapping from DFT to TB is a simple projection. A future study may unveil whether other projectors are better.
%The important point is to define the TB to be computationally manageable, along with a coupling between the pristine TB and DFT regions, such that the interface scattering is much smaller than the scattering mechanisms under study.
%
We further anticipate that the basic concepts of this method can be generalized to investigate thermal transport, where multi-scale approaches are often crucial.\cite{Mortazavi2017, Mortazavi2014, Nozaki2011} For this purpose the open-source \phtrans\ package\cite{PapiorThesis} (a \tbtrans\ variant) may turn out to be very helpful.

\section*{Acknowledgments}
Financial support by Villum Fonden (00013340), Danish research council (4184-00030) and EU-Horizon 2020 research and innovation programme (766726) is gratefully acknowledged. The Center for Nanostructured Graphene (CNG) is sponsored by the Danish Research Foundation (DNRF103). 

%%%END OF MAIN TEXT%%%

%%%%%%%%%%%%%%%%%%%%%
\section{Supplementary Material}
(Available online) Large-scale TB transport calculations of a $60\times35 \,\nm^2$ graphene device with N=3 DFT-precision regions, namely one epoxy defect and two source and drain Au tips, both in contact to graphene $\approx 2\,\angstrom$ above an hexagon. Two regular graphene electrodes are used along $y$ and CAP is set at the outermost $5\,\nm$ on cell boundaries along $x$ to absorb currents.
The file shows bond-currents injected by one of the two tips into the system, scattering off the drain tip and the epoxy defect, for different positions of the epoxy defect with respect to the two tips.

%%%REFERENCES%%%
\section*{References}
%\bibliography{multiscale_bib_mb} 
\bibliographystyle{unsrt}

\end{document}